\documentclass[11pt]{article}

\usepackage{graphics,wrapfig,times}
\usepackage{graphicx}

\usepackage{epsfig,epstopdf}
\usepackage{amsfonts}
\usepackage{mathrsfs}
\usepackage{amsmath,amssymb}
\usepackage[T1]{fontenc}

\begin{document}

\title{An Optimal Choice of Reference for the Quasi-Local Gravitational Energy and Angular Momentum \\
$ $\\
{\small Gang Sun$^1$, Chiang-Mei Chen$^{1,2}$, Jian-Liang Liu$^{1}$, James M. Nester$^{1,2,3,4}$ \\
$ $\\
{$^{1}$Department of Physics, National Central University, Chungli, 320, Taiwan, \\
$^{2}$Center for Mathematics and Theoretical Physics, \\
National Central University, Chungli 320, Taiwan \\
$^{3}$Graduate Institute of Astronomy, National Central University, Chungli 320, Taiwan \\
$^{4}$Institute of Physics, Academia Sinica, Taipei 115, Taiwan \\
liquideal@hotmail.com, cmchen@phy.ncu.edu.tw, liujl@phy.ncu.edu.tw, nester@phy.ncu.edu.tw}}}

\maketitle

\begin{abstract}
The boundary term of the gravitational Hamiltonian can be used to give the values of the quasi-local quantities as long as one can provide a suitable evolution vector field and an appropriate reference. On the two-surface boundary of a region we have proposed using {\em four dimensional isometric matching\/} between the dynamic spacetime and the reference geometry along with energy extremization  to find both the optimal reference matching and the appropriate quasi-Killing vectors. Here we  consider the axisymmetric spacetime case.  For the Kerr metric in particular we can explicitly solve the equations to find the best matched reference and quasi-Killing vectors.  This leads to the exact expression for the quasi-local boundary term  and the values of our optimal quasi-local energy and angular momentum.\\


{\it Pacs}: {04.20.Cv, 04.20.Fy}
\end{abstract}

%\maketitle

%%%%%%%%%%%%%%%%%%%%%%%%%%%%%%%%%%%%%%%%%%%%%%%%%%%%%%%%%%%%%%%%%%%%%%
\section{Introduction}
%%%%%%%%%%%%%%%%%%%%%%%%%%%%%%%%%%%%%%%%%%%%%%%%%%%%%%%%%%%%%%%%%%%%%%

People have been concerned with how to define the energy-momentum and angular momentum of gravitating systems since even before general relativity was born. Different pseudotensors were proposed to describe the local gravitational energy-momentum density. This choice between the different available expressions is one  ambiguity in characterizing the energy-momentum of any local region of space-time. Moreover there is an additional ambiguity: since a pseudotensor is not a covariant object, it leads to values which inherently depend on the reference frame. This feature is actually an inevitable consequence of the fundamental geometric nature of gravity. Mathematically it is due to diffeomorphism invariance (as shown by Noether in her famous paper~\cite{noether}); physically it has been said that the equivalence principle forbids the localization of gravitational energy~\cite{MTW73}. The more modern view is that energy-momentum should be regarded as being \emph{quasi-local} (associated with a closed 2-surface) rather than local. The various quasi-local ideas have been comprehensively reviewed~\cite{Sza09}. There have been many quasi-local proposals, but no consensus.  In fact the quasi-local approach has ambiguities analogous to the two already mentioned.  However, it has been argued that the Hamiltonian approach, in particular the covariant Hamiltonian formalism~\cite{N91,N95,N99,N00,Nester04,N05,N08}, can tame both of these ambiguities, clarifying their geometric and physical nature.    The quasi-local values are then determined by the Hamiltonian boundary term.  A preferred boundary term has been identified; this gives a resolution to the first type of ambiguity. Recent work has finally led to identifying a good procedure for addressing the second type of ambiguity: how to best select the necessary reference values on the boundary~\cite{ae100}. We now test this program on axisymmetric spacetimes, showing that it works well; in particular it yields sensible energy and angular momentum values for the Kerr metric.

Early quasi-local investigations (an outstanding one is the seminal work of Brown and York~\cite{BY93}) naturally explicitly worked out quasi-local energy values for spherically symmetric spacetimes.  As far as we know Matrinez~\cite{Mart94} was the first to consider an axisymmetrical spacetime.  We are concerned with the covariant Hamiltonian formalism; within this approach the quasi-local energy for spherically symmetric spacetime was examined in depth by Wu {\it et al}.~\cite{Wu10,Wu11,Liu11,Wu12}  Here we will now go further: using the results of the covariant Hamiltonian formalism we consider an axisymmetric spacetime, use a new approach to find the best matched reference, and determine both the optimal quasi-local energy and angular momentum.

%%%%%%%%%%%%%%%%%%%%%%%%%%%%%%%%%%%%%%%%%%%%%%%%%%%%%%%%%%%%%%%%%%%%%%
\section{Covariant Hamiltonian Formalism}
%%%%%%%%%%%%%%%%%%%%%%%%%%%%%%%%%%%%%%%%%%%%%%%%%%%%%%%%%%%%%%%%%%%%%%

Detailed discussions of the covariant Hamiltonian formalism as developed by our research group can be found in the references~\cite{N91,N95,N99,N00,Nester04,N05,N08}. Here we give a brief summary of some key features.

A \emph{first order Lagrangian 4-form} for a $k$-form field $\phi$ has the form $\mathcal{L}=d\phi\wedge p-\Lambda(\phi,p)$. A vector field $N$ can be regarded as an ``infinitesimal translation''. It generates a local infinitesimal diffeomorphism under which the action associated with $\mathcal{L}$ should be invariant. Using the well-known Lie derivative formula $\pounds_N \equiv di_N + i_Nd$, the associated conserved\footnote{Conserved in the first Noether theorem sense: $d{\cal H}(N) \propto $ field equations.} Noether current 3-form---which is moreover the Hamiltonian---
\begin{equation}
\mathcal{H}(N) := \pounds_{N} \phi \wedge p - i_{N} \mathcal{L} \equiv N^{\mu} \mathcal{H}_{\mu} + d\mathcal{B}(N),
\end{equation}
can be split as indicated into two pieces with distinct roles. The part algebraic in the vector field generates the Hamiltonian equations; from Noether's second theorem concerning local invariance with respect to the choice of $N$ one finds that it is proportional to field equations and thus vanishes ``on shell''. Then the \emph{value} of the Hamiltonian,
\begin{eqnarray} \label{eq1}
E(N, \Sigma) = H(N, \Sigma) = \int_{\Sigma} \mathcal{H}(N) = \oint_{\partial\Sigma} \mathcal{B}(N),
\end{eqnarray}
is {\em quasi-local\/}; it is determined entirely by quantities on the closed 2-surface $S := \partial\Sigma$.

For \emph{any} choice of $N$ this expression defines a conserved quasi-local quantity.  To get a physically meaningful result one needs to have a good way of selecting the specific vectors $N$ that correspond to the quasi-local symmetries that yield the quasi-local energy-momentum and angular momentum/center-of-mass.

It must be noted that the Hamiltonian boundary 2-form ${\cal B}(N)$ can be modified ``by hand''; this would still preserve the conservation property, but change the conserved value. It was shown that in this way the Hamiltonian approach can include the various pseudotensors~\cite{Chang99}. But this ``Noether current'' ambiguity is tamed by the Hamiltonian \emph{boundary variation principle\/}: the boundary term in the variation of the Hamiltonian must be required to vanish.  This determines the related boundary conditions: different boundary terms are associated with different boundary conditions.

Here we consider specifically only Einstein's theory, general relativity (GR).
%The basic dynamical variables used are the orthonormal coframes $\vartheta^{\alpha}=\vartheta^{\alpha}_{~k}dx^{k}$
%and the connection one-forms $\omega^{\alpha}{}_{\beta}=\Gamma^{\alpha}{}_{\beta k}dx^{k}$.
%
Our preferred\footnote{At spatial infinity it gives the accepted energy, momentum, angular momentum, and center-of-mass. At null infinity it gives the Bondi energy and energy flux, for small spheres it is a positive multiple of the Bel-Robinson tensor, it has a positivity property, and for spherically symmetric solutions it has the hoop property~\cite{Omur10}.}
boundary term for GR (which corresponds to holding the metric fixed on the boundary) is
\begin{eqnarray}\label{eq2}
16 \pi \mathcal{B}(N) = \Delta\Gamma^\alpha{}_\beta \wedge i_{N} \eta_{\alpha}{}^{\beta} + \overline{D}_{\beta} N^{\alpha} \Delta\eta_{\alpha}^{~\beta}, \label{B}
\end{eqnarray}
where $\eta^{\alpha\beta} := *(dx^\alpha\wedge dx^\beta)$ is the dual form basis, $\Gamma^\alpha{}_\beta = \Gamma^\alpha{}_{\beta\mu}dx^\mu$ is the connection one form, and $\Delta\Gamma := \Gamma - \bar\Gamma$, $\Delta\eta := \eta - \bar\eta$, with the bar refering to non-dynamical reference values
(which are needed only in a neighborhood of the boundary $S$).
The component form of our expression\footnote{An equivalent expression was proposed by Katz, Bi\v{c}\'ak, and Lynden-Bel~\cite{LBKB95,Katz:1996nr}.} is
\begin{eqnarray}
16 \pi \mathcal{B}(N) \label{eq3}
&=& \frac{1}{4} [ N^{\sigma} \sqrt{-g} g^{\gamma\beta} \Delta\Gamma^{\alpha}{}_{\beta\lambda} \delta^{\lambda\tau\rho}_{\alpha\gamma\sigma}
\nonumber\\
&& \quad + \overline{D}_{\beta} N^{\alpha} \Delta(\sqrt{-g} g^{\gamma\beta}) \delta^{\tau\rho}_{\alpha\gamma} ] \epsilon_{\tau\rho\mu\nu} dx^{\mu} \wedge dx^{\nu}.
\end{eqnarray}

The reference values are non-dynamic, they represent the ``vacuum'' or ground state; how to best select these values is our concern here.
The natural reference for dynamic geometry is Minkowski space.  Thus we need to select a specific Minkowski space in the neighborhood of the two boundary.  Any such choice can be specified by selecting some \emph{quasi-Minkowski coordinates}; in fact any four independent functions $y^i,\ i = 0 \dots 3$ defined in a neighborhood of the the 2-boundary $S$ will define a Minkowski metric by $d\bar s^2 = \bar g_{ij} dy^i dy^j, \ \bar g_{ij} = \hbox{diag}(-1,1,1,1)$.
A Killing field of this reference has the form $N^i = N^i_0 + \lambda_0^i{}_j y^j$, where the translation parameters $N_0^i$ and the boost-rotation parameters $\lambda_0^{ij} = \lambda_0^{[ij]}$ are constants.
Using this reference the 2-surface integral  of the Hamiltonian boundary term then gives a value of the form
\begin{equation}
\oint_S {\mathcal B}(N) = - N_0^i p_i(S) + \frac12 \lambda^0_{ij} J^{ij}(S)\,,
\end{equation}
which yields not only a quasi-local \emph{energy-momentum} but also a quasi-local \emph{angular momentum/center-of-mass}. These quasi-local values of course depend on the chosen reference.  As long as the reference approaches at an appropriate rate the flat Minkowski space at spatial infinity, the integrals $p_i(S),\ J^{ij}(S)$ in the spatial asymptotic limit will agree with accepted expressions for these quantities~\cite{MTW73,Regge:1974zd,Beig:1987zz,Szabados:2003yn,Sza06}.

%%%%%%%%%%%%%%%%%%%%%%%%%%%%%%%%%%%%%%%%%%%%%%%%%%%%%%%%%%%%%%%%%%%%%%
\section{Optimal choice for reference}
%%%%%%%%%%%%%%%%%%%%%%%%%%%%%%%%%%%%%%%%%%%%%%%%%%%%%%%%%%%%%%%%%%%%%%

To explicitly determine the specific values of the quasi-local quantities one needs some good way to choose the reference.  Minkowski spacetime is the natural choice, especially for asymptotically flat spacetimes~\cite{Kij97}.  However, as noted above, almost any four functions will determine some reference. With such freedom one can still get almost any value for the quasi-local quantities.  This freedom is the quasi-local version of the second type of ambiguity mentioned in the introduction in connection with the reference frame dependence of energy-momentum pseudotensor expressions.

Recently we have proposed a procedure~\cite{ae100} to fix the ``best'' choice of reference.  It has two features: 4D isometric matching\footnote{This was already proposed by Szabados back in 2000 (at a workshop in Hsinchu, Taiwan); he has since extensively explored this idea~\cite{Sza05}.} and energy optimization.  Here we shall show that our procedure works well for a certain class of axisymmetric spacetimes.

\subsection{4D isometric matching}

Previously the program of 4-dimensional isometric matching with energy extremization was explicitly worked out for the most important special case: spherically symmetric spaces (both static and dynamic~\cite{Wu11,Wu12}). Here we turn to what is probably the next most important special case: {\em axisymmetry}. More specifically we will consider Kerr-like metrics, with just one non-vanishing off diagonal term. (The most general axisymmetric metric and the general case without any symmetry are also being investigated; the results of these investigations will be presented in due course.)
Specifically, in terms of suitable spherical coordinates
$x^\mu = \{ x^{0}, x^{1}, x^{2}, x^{3} \} = \{ t, r, \theta, \varphi\}$, we consider metrics of the form %and the spacetime metric with only one cross component.
\begin{eqnarray}\label{eq20}
ds^2 = g_{00} dt^2 + g_{11} dr^2 + g_{22} d\theta^2 + g_{33} d\varphi^2 + 2 g_{03} dt d\varphi, \label{metric}
\end{eqnarray}
where the metric components are independent of $\varphi$.
Although it is not the most general metric of an axisymmetric spacetime, it is enough to unveil the property of a non-static spacetime with angular momentum so we can test our ideas.

We assume that the spherical-like coordinates can be chosen such that the boundary of 2-sphere $S = \partial\Sigma$ is given by $t=t_{0}, r=r_{0}$. In general the reference Minkowski metric is determined by four quasi-Minkowski coordinates defined in a neighborhood of $S$ by four suitable functions, $y^i = (T, X, Y, Z)$ of $(t, r, \theta, \varphi)$; from the $y^i$ we find $dy^i = y^i{}_\mu dx^\mu$.

Since the dynamic spacetime metric is not diagonal when expanded in terms of $x^\mu$ the reference metric must likewise allow for off diagonal terms.
In the axisymmetric case one convenient way to achieve this is to use a certain cylindrical form. Let
\begin{eqnarray}\label{eq21}
X = \rho \cos(\varphi + \Phi), \qquad Y = \rho \sin(\varphi + \Phi),
\end{eqnarray}
and assume that $(T, \rho, \Phi, Z)$ are axisymmetric, i.e., they do not depend on $\varphi$.
Then the reference Minkowski metric in a neighborhood of $S$ can be taken to have the form
\begin{eqnarray}\label{eq22}
d \bar{s}^2 &=& \bar g_{ij} dy^i dy^j
\nonumber\\
&=& - dT^2 + dX^2 + dY^2 + dZ^2 = - dT^2 + d\rho^2 + \rho^2 (d\varphi + d\Phi)^2 + dZ^2
\nonumber\\
&=& - ( T_{t} dt + T_{r} dr + T_{\theta} d\theta )^2 + ( \rho_{t} dt + \rho_{r} dr + \rho_{\theta} d\theta )^2
\nonumber\\
&& + \rho^2 ( d\varphi + \Phi_{t} dt + \Phi_{r} dr + \Phi_{\theta} d\theta )^2 + ( Z_{t} dt + Z_{r} dr + Z_{\theta} d\theta )^2
\\
&=& \bar g_{ij} y^i{}_\mu y^j{}_\nu dx^\mu dx^\nu = \bar g_{\mu\nu} dx^\mu dx^\nu. \nonumber
\end{eqnarray}
On the 2-sphere $S$ determined by $t = t_0, r = r_0$ the reference metric is a function of 12 functions of $\theta$, namely\footnote{In this work, we will just deal with a special type of metric with only one cross component. However, (\ref{eq21}), (\ref{eq22}),  and (\ref{isometric})--(\ref{g33}) are quite general, so we can straightforwardly extend our procedure to cover the most general axisymmetric metric.}
\begin{eqnarray}
\{ T_{t}, T_{r}, T_{\theta}, \rho_{t}, \rho_{r}, \rho, \Phi_{t}, \Phi_{r}, \Phi_{\theta}, Z_{t}, Z_{r}, Z_{\theta} \}
\end{eqnarray}
(note that on the 2-sphere $\rho$ determines $\rho_\theta := \partial_\theta\rho$).  We take $T_{r}$ and $T_{\theta}$ as a suitable set of controlling functions on the 2-sphere; then the other 10 can be found algebraically from the 4D isometric matching conditions, $ds^2 \doteq d\bar s^2$, i.e.,
\begin{equation}
g_{\mu\nu} \doteq \bar g_{\mu\nu} := \bar g_{ij} y^i{}_\mu y^j{}_\nu. \label{isometric}
\end{equation}
(Here the notation $\doteq$ means that the relation holds just on the 2-surface $S$.)
For a general metric these relations have the detailed form
\begin{eqnarray}
g_{00} &\doteq& -T_{t}^2 + \rho_{t}^2 + \rho^2 \Phi_{t}^2 + Z_{t}^2, \label{g00} \\
g_{01} &\doteq& -T_{t} T_{r} + \rho_{t} \rho_{r} + \rho^2 \Phi_{t} \Phi_{r} + Z_{t} Z_{r}, \label{g01} \\
g_{11} &\doteq& -T_{r}^2 + \rho_{r}^2 + \rho^2 \Phi_{r}^2 + Z_{r}^2, \label{g11} \\
g_{02} &\doteq& -T_{t} T_{\theta} + \rho_{t} \rho_{\theta} + \rho^2 \Phi_{t} \Phi_{\theta} + Z_{t} Z_{\theta}, \label{g02} \\
g_{03} &\doteq& \rho^2 \Phi_{t}, \label{g03} \\
g_{12} &\doteq& -T_{r} T_{\theta} + \rho_{r} \rho_{\theta} + \rho^2 \Phi_{r} \Phi_{\theta} + Z_{r} Z_{\theta}, \label{g12} \\
g_{13} &\doteq& \rho^2 \Phi_{r}, \label{g13} \\
g_{22} &\doteq& -T_{\theta}^{2} + \rho_{\theta}^{2} + \rho^{2} \Phi_{\theta}^{2} + Z_{\theta}^{2}, \label{g22} \\
g_{23} &\doteq& \rho^2 \Phi_{\theta}, \label{g23} \\
g_{33} &\doteq& \rho^2. \label{g33}
\end{eqnarray}

For our special axisymmetric metric with only one cross term it is easy to see from (\ref{g03}, \ref{g13}, \ref{g23}, \ref{g33}) that
\begin{equation}
\rho \doteq \sqrt{g_{33}}, \quad \Phi_{t} \doteq {g_{03}}/{g_{33}}, \quad \Phi_{r} \doteq g_{13}/g_{33} = 0, \quad \Phi_{\theta} \doteq g_{23}/g_{33} = 0.
\end{equation}
Recalling that $\rho_\theta$ follows from $\rho$, from (\ref{g22}) one can now get\footnote{Consider the spherical coordinate system in $\mathbf{R}^3$.
The relation between the Cartesian coordinates $X, Y, Z$ and the spherical coordinates $r, \theta, \varphi$ is $X = r \sin\theta \cos\varphi$, $Y = r \sin\theta \sin\varphi$, $Z = r \cos\theta$, which is such that $Z_{\theta} = -r \sin\theta$.  Hence the negative root is the reasonable one here, for we wish our procedure to give the usual result for metrics which are flat.}
%Below we will find an analogous situation for $\rho_{r}$ and $T_{t}$.
%Then one can similarly consider their analogues in a three dimensional coordinate system and infer the appropriate sign.}
%
\begin{equation}
\nonumber Z_{\theta} \doteq - \sqrt{g_{22} + T_{\theta}^2 - \rho_{\theta}^2}.
\end{equation}
This leaves the remaining unknowns $\{ T_{t}, \rho_{t}, Z_{t}, \rho_{r}, Z_{r} \}$ to be determined.
%via 2 linear equations (\ref{}and 3 quadratic equations. Thus,
Consider next the pair (\ref{g11}, \ref{g12}):
\begin{eqnarray}
\rho_{r} \rho_{\theta} + Z_{r} Z_{\theta} &\doteq& T_{r} T_{\theta} + g_{12} = T_{r} T_{\theta},
\\
\rho_{r}^2 + Z_{r}^2 &\doteq& g_{11} + T_{r}^2.
\end{eqnarray}
One can first get $Z_{r}$,
\begin{eqnarray}
Z_{r} \doteq \frac{T_{r} T_{\theta} - \rho_{r} \rho_{\theta}}{Z_{\theta}}, \label{Zr}
\end{eqnarray}
and then a quadratic equation for $\rho_{r}$, namely
\begin{eqnarray}
\rho_{r}^2 \left( 1 + \frac{\rho_{\theta}^2}{Z_{\theta}^2} \right) - 2 \frac{\rho_{r} \rho_{\theta} T_{r} T_{\theta}}{Z_{\theta}^2} + \frac{T_{r}^2 T_{\theta}^2}{Z_{\theta}^2} - T_{r}^2 - g_{11} \doteq 0\,.
\end{eqnarray}
Generally $\rho_{r}$ has two solutions,
but one should select the sign that gives a positive value---at least when the dynamical geometry is near flat space.
Once $\rho_{r}$ is determined, $Z_{r}$ can be found from (\ref{Zr}).

As to $T_{t}, \rho_{t}, Z_{t}$, they can now be found through the remaining relations (\ref{g00}, \ref{g01}, \ref{g02}) which are effectively one quadratic and two linear equations:
\begin{eqnarray}
-T_{t}^2 + \rho_{t}^2 + Z_{t}^2 &\doteq& g_{00} - \rho^2 \Phi_{t}^2, \label{00}
\\
-T_{t} T_{r} + \rho_{t} \rho_{r} + Z_{t} Z_{r} &\doteq& g_{01} = 0,
\\
-T_{t} T_{\theta} + \rho_{t} \rho_{\theta} + Z_{t} Z_{\theta} &\doteq& g_{02} = 0.
\end{eqnarray}
From the two linear equations we find
\begin{eqnarray}
\rho_{t} &\doteq& T_{t} \, \frac{T_{r} Z_{\theta} - T_{\theta} Z_{r}}{\rho_{r} Z_{\theta} - \rho_{\theta} Z_{r}}, \label{rhot}
\\
Z_{t} &\doteq& T_t\, \frac{\rho_r T_{\theta} - T_r \rho_{\theta}}{\rho_{r} Z_{\theta} - \rho_{\theta} Z_{r}}. \label{Zt}
\end{eqnarray}
Substituting this into (\ref{00}) gives
%Graphically the system is a line intersecting a hyperbola so that two solutions are expected.
a new quadratic equation for $T_{t}$:
\begin{eqnarray}
T_{t}^2 \left[ -1 + \frac{(T_{r} Z_{\theta} - T_{\theta} Z_{r})^2}{(\rho_{r} Z_{\theta} - \rho_{\theta} Z_{r})^2} + \frac{(\rho_{r} T_{\theta} - \rho_{\theta} T_{r})^2}{(\rho_{r} Z_{\theta} - \rho_{\theta} Z_{r})^2} \right] \doteq g_{00} - \rho^2 \Phi_{t}^2.
\end{eqnarray}
This has two solutions, but one should choose the positive value.
Once $T_{t}$ is found, $\rho_{t}$ and $Z_{t}$ can be found from (\ref{rhot}, \ref{Zt}).

Algebraically solving these equations and collecting our isometric matching results together, we have in addition to vanishing $\Phi_r,\ \Phi_\theta$,
\begin{eqnarray}
\rho &\doteq& \sqrt{g_{33}}, \\
\Phi_{t} &\doteq& g_{03}/g_{33}, \\
%\quad \Phi_{r} \doteq 0 \doteq \Phi_{\theta}, \\
Z_{\theta} &\doteq& - L_{1}, \\
T_{t} &\doteq& L_{2} L_{3}, \\
\rho_{r} &\doteq& (\rho_{\theta} T_{r} T_{\theta} + {L_{1} L_{3}})/{L_{4}}, \\
Z_{r} &\doteq& ({\rho_{\theta} L_3} - {L_{1} T_{r} T_{\theta}})/{L_{4}}, \\
\rho_{t} &\doteq& L_2 ({L_{1} T_{r} g_{22}} + {L_3 T_{\theta} \rho_{\theta}})/{L_{4}}, \\
Z_{t} &\doteq& L_2 ({\rho_{\theta} T_{r} g_{22}} - {L_{1} L_3 T_{\theta}})/{L_{4}},
\end{eqnarray}
where
\begin{eqnarray}
L_{1} &:=& \sqrt{g_{22} - \rho_{\theta}^2 + T_{\theta}^2}, \\
L_{2} &:=& \sqrt{\frac{g_{03}^2 - g_{00} g_{33}}{g_{11} g_{22} g_{33}}}, \\
L_{3} &:=& \sqrt{T_{r}^2 g_{22} + g_{11} g_{22} + g_{11} T_{\theta}^2}, \\
L_{4} &:=& g_{22} + T_{\theta}^2.
\end{eqnarray}

With 4D isometric matching on the 2-boundary $S$ the second term in our quasi-local expression (\ref{B}) vanishes, so it simplifies considerably to
\begin{equation}
16 \pi \mathcal{B}(N) = (\Gamma^\alpha{}_\beta - \bar\Gamma^\alpha{}_\beta) \wedge i_{N} \eta_{\alpha}{}^{\beta}. \label{B2}
\end{equation}
We will use this expression to determine our quasi-local energy and angular momentum.

%The Killing fields of the reference give the \emph{quasi-symmetries} of the 2-surface $S$.\cite{Kij02}
%
%  Choosing the vector field $N$ to be the various Killing fields of the reference gives the associated  quasi-local quantities.
%  In this program $N$ is a function of the embedding variables.

\subsection{Energy extremization}
Even with 4D isometric matching the quasi-local values still have a lot of freedom. The value of the energy-momentum can be regarded as a measure of the difference between the dynamical geometry and the reference geometry.  With this in mind it was proposed to look to the critical points of $m^2 := - \bar g^{ij} p_i p_j$~\cite{ae100}.  Here we shall use a simpler expression which should lead to the same answer.
To properly select  $T_{r}$ and $T_{\theta}$ we propose extremizing the quasi-local energy to find the best matching for these embedding functions.

The quasi-local energy is determined by selecting the reference geometry unit timelike Killing vector\footnote{The subscript in $N_{\rm E}$ is used
to denote the vector field for energy}: $N_{\rm E} = \partial_{T}$.
To calculate the quasi-local energy one needs  the components of $N_{\rm E}$ and $\bar\Gamma^\alpha{}_\beta$ in the dynamic spacetime.
The reference connection is determined by pullback from the flat space reference (which vanishes in Minkowski coordinates):
\begin{equation}
\bar \Gamma^\alpha{}_\beta = x^\alpha{}_i (\bar\Gamma^i_{~j} y^j{}_\beta + dy^i{}_\beta) = x^\alpha{}_i dy^i{}_\beta.
\end{equation}
Here $x^\alpha{}_i$ is the inverse of $y^i{}_\alpha$, and the latter is determined from $dy^i = y^i{}_\alpha dx^\alpha$, i.e.,
\begin{eqnarray} \label{reference choice}
\left(
  \begin{array}{c}
    dT \\
    dX \\
    dY \\
    dZ \\
  \end{array}
\right)=\left(
          \begin{array}{cccc}
            T_t & T_r & T_\theta & 0 \\
            \rho_t \cos\xi - \rho \sin\xi \Phi_t\ \ & \rho_r \cos\xi\ \ & \rho_\theta \cos\xi\ \ & -\rho \sin\xi \\
            \rho_t \sin\xi + \rho \cos\xi \Phi_t\ \ & \rho_r \sin\xi\ \ & \rho_\theta \sin\xi\ \ & \rho \cos\xi \\
            Z_t & Z_r & Z_\theta & 0 \\
          \end{array}
        \right)\left(
                 \begin{array}{c}
                   dt \\
                   dr \\
                   d\theta \\
                   d\varphi \\
                 \end{array}
               \right),
\end{eqnarray}
where $\xi = \varphi + \Phi$.

Using these results, the determined values for the components of the vector are
\begin{eqnarray}\label{eq46}
&& N^{t}_{\rm E} \doteq \frac{L_3}{g_{11} g_{22} L_2}, \qquad N^{r}_{\rm E} \doteq - \frac{T_{r}}{g_{11}}, \qquad N^{\theta}_{\rm E} \doteq -\frac{T_{\theta}}{g_{22}},
\nonumber\\ %&\qquad&
&& N^{\varphi}_{\rm E} \doteq - \Phi_t N^t_{\rm E} = - \frac{g_{03} L_3}{g_{11} g_{22} g_{33} L_2}. \label{N}
\end{eqnarray}
With $N = N_{\rm E}$ the quasi-local boundary term (\ref{B2}) works out to be
\begin{eqnarray}
16 \pi \mathcal{B}(N_{\rm E}) \!&\!\!=\!\!&\! \sqrt{g_{33}} \biggl\{ \frac{2 T_{\theta}}{L_{3}} T_{r\theta} + \frac{2}{L_4} \left[ -\frac{T_{\theta}^2 T_{r}}{L_{3}} + \frac{T_{\theta} \rho_{\theta}}{L_{1}} \right] T_{\theta\theta}
- \frac{g_{11\theta} T_r T_\theta}{L_3 g_{11}}
\nonumber\\
&& \quad + \frac{g_{22\theta}}{L_4} \left[ \frac{T_r T_\theta^3}{g_{22} L_3} + \frac{\rho_\theta}{L_1} \right]
%\\
- \frac{2 \rho_{\theta\theta}}{L_{1}} + \frac{2 L_{1}}{\sqrt{g_{33}}} + L_{3} W \biggr\} d\theta \wedge d\varphi, \qquad \label{BE}
\end{eqnarray}
with
\begin{equation}
W := -\frac{1}{g_{11}} \left( \frac{g_{22r}}{g_{22}} + \frac{g_{33r}}{g_{33}} \right).
\end{equation}
Here $g_{22\theta}$ denotes its derivative by $\theta$ and so on.  Also it should be noted that in this expression and in all of the following, to convey our ideas without too many complications, we have additionally assumed that the  metric components in (\ref{metric}) are all time independent.

The quasi-local energy value is the integral of the 2-form $\mathcal{B}(\partial_T)$ (\ref{BE}) over $S$.  The value is a functional of the two control functions $T_\theta, T_r$.  The distinguished values of this energy functional are the ``critical points'', which can be found as the solutions of  the respective variational derivatives, the two Euler-Lagrange equations obtained by variation with respect to the control variables $T_\theta,T_r$.  The respective variational equations are
\begin{eqnarray}\label{eq50}
0 &=& \frac{\partial \mathcal{B}(\partial_T)}{\partial T_{\theta}} - \frac{d}{d\theta} \frac{\partial \mathcal{B}(\partial_T)}{\partial T_{\theta\theta}}
\nonumber\\
&=& \frac{\sqrt{g_{33}}}{L_3} \left[ 2 T_{r\theta} + \left( -\frac{g_{11\theta}}{g_{11}} + \frac{g_{33\theta} T_{\theta}^2}{g_{33} L_{4}}\right) T_{r} + \frac{2 L_{1} L_3 T_{\theta}}{\sqrt{g_{33}} L_{4}} + g_{11} W T_{\theta} \right], \label{ELtheta}
\\
\label{eq51}
0 &=& \frac{\partial \mathcal{B}(\partial_T)}{\partial T_{r}} - \frac{d}{d\theta} \frac{\partial \mathcal{B}(\partial_T)}{\partial T_{r\theta}}
\nonumber\\
&=& \frac{\sqrt{g_{33}}}{L_3} \left[ -2 T_{\theta\theta} + \left( \frac{g_{22\theta}}{g_{22}} - \frac{g_{33\theta}}{g_{33}} \right) T_{\theta} + g_{22} W T_{r} \right]. \label{ELr}
\end{eqnarray}
The second relation can be used to eliminate $T_r$ from the first relation.  The result is a quasi-linear (but highly nonlinear) second order equation for $T_\theta$.  For such equations there is generally little hope of finding any analytic solution.  However, this case is still rather special: there is one quite obvious simple solution that applies to all of these Kerr-like metrics, namely the trivial solution $T_r = 0 = T_\theta$.

For the simple solution $T_r=0=T_\theta$ the vector field takes the simpler form
\begin{eqnarray}\label{eq58}
N^{t}_{\rm E} = \frac{1}{T_{t}} = \sqrt{\frac{g_{33}}{g_{03}^2 - g_{00}g_{33}}}, \quad N^{\varphi}_{\rm E} = -\frac{g_{03}}{g_{33}} N^t_{\rm E}, \quad N^{r}_{\rm E} = 0, \quad N^{\theta}_{\rm E} = 0.\label{vector}
\end{eqnarray}
This vector is orthogonal to the constant $t$ hypersurface; physically it can be identified with the \emph{static\/} observer on the boundary.

From (\ref{BE}) with $T_\theta=0=T_r$, on the $t,r$ constant 2-surface of any axisymmetric time independent metric of the form (\ref{eq20}), our optimal boundary term for the quasi-local energy is
\begin{eqnarray}\label{eq4}
\mathcal{B}_{\rm E} &=& \frac{\sqrt{g_{33}}}{16\pi} \Bigl\{ -\Bigl( \frac{\partial_{r} g_{22}}{g_{22}} + \frac{\partial_{r} g_{33}}{g_{33}} \Bigr) \frac{\sqrt{g_{22}}}{\sqrt{g_{11}}}
\nonumber\\
&+& \frac{(\partial_{\theta} g_{22})(\partial_{\theta} g_{33}) - 2 g_{22} (\partial_{\theta\theta} g_{33}) + 4 g_{22}^2}{g_{22} \sqrt{4 g_{33} g_{22} - (\partial_{\theta} g_{33})^2}} ~\Bigr\} d\theta \wedge d\varphi. \label{optimalenergy}
\end{eqnarray}

Turning to the quasi-local angular momentum, note that it can have only one non-vanishing component.  To select this component from our quasi-local boundary term (\ref{B2}) the appropriate vector field is the \emph{reference rotational Killing field}, $N_{\rm AM} = X \partial_{Y} - Y \partial_{X}$, around the axis. In the dynamic spacetime, this vector field works out to have a very simple form, namely
\begin{eqnarray}\label{eq54}
N_{\rm AM} = \partial_{\varphi},
\end{eqnarray}
and it is in fact the \emph{axisymmetric Killing field} for the metrics considered.
%Liu_130618
For the choice of $N_{\rm AM} = \partial_{\varphi}$, which implies that the only non-vanishing vector component is $N^{\varphi} = 1$, the boundary expression (\ref{B2}) takes the specific form
\begin{eqnarray}\label{AM N3=1}
16 \pi \mathcal{B}_{\rm AM} = \sqrt{-g} (g^{1\beta} \Delta\Gamma^{0}{}_{\beta3} - g^{0\beta} \Delta\Gamma^{1}{}_{\beta 3} ) d\theta \wedge d\varphi.
\end{eqnarray}
Using the reference choice (\ref{reference choice}), a computation shows that the reference connection contribution vanishes, so the boundary term depends only on the dynamical part:
\begin{eqnarray}\label{AM N3=1 dyn}
16 \pi \mathcal{B}_{\rm AM} &=& \sqrt{-g} (g^{11} \Gamma^{0}{}_{13} - g^{00} \Gamma^{1}{}_{03} - g^{03} \Gamma^{1}{}_{33}) d\theta \wedge d\varphi
\nonumber\\
&=& \sqrt{-g} g^{11} (g^{00} \partial_{r} g_{03} + g^{03} \partial_{r} g_{33}) d\theta \wedge d\varphi. \label{J}
%&=&2ma\sin^{3}\theta[(1-2r^{2}/\Sigma)(r^{2}+a^{2})-2r^{2}/\Sigma]d\theta\wedge d\varphi.
\end{eqnarray}
%Liu_130618

%Briefly, we  found that the physically meaningful solution of (\ref{eq50}) and (\ref{eq51}) occurs at $T_{r}=-\lambda\cos\theta,~T_{\theta}=\lambda r\sin\theta$ with a constant $\lambda$ for the Minkowski metric, and at $T_{r}=T_{\theta}=0$ for the Schwarzschild and the Kerr metric, the details are presented in the next section.  We found that in these cases the critical points always correspond to the maximum energy value.

%%%%%%%%%%%%%%%%%%%%%%%%%%%%%%%%%%%%%%%%%%%%%%%%%%%%%%%%%%%%%%%%%%%%%%
\section{Detailed results}
%%%%%%%%%%%%%%%%%%%%%%%%%%%%%%%%%%%%%%%%%%%%%%%%%%%%%%%%%%%%%%%%%%%%%%

To see whether this simple choice is really a good choice---maybe even the best---for determining the reference we examined some physically meaningful special cases.

%*********************************************************************

\subsection{Minkowski metric}
For the Minkowski metric it is easy to see that there is actually a simple non-trivial solution to the two embedding equations (\ref{eq50}) and (\ref{eq51}), namely
\begin{equation}
T_{r} = -\lambda \cos\theta, \qquad T_{\theta} = \lambda r \sin\theta , \label{boostedref}
\end{equation}
with a constant $\lambda$. With this choice the vector field (\ref{N}) corresponds to that of a constant boost along the axis:
\begin{eqnarray}\label{eq53}
N^{t}_{\rm E} = \sqrt{\lambda^2 + 1}, \quad N^{r}_{\rm E} = \lambda \cos\theta, \quad N^{\theta}_{\rm E} = -\frac{\lambda\sin\theta}{r}, \quad N^{\varphi}_{\rm E} = 0.
\end{eqnarray}
For all $\lambda$ the value of the optimal boundary expression (\ref{BE}) and the associated energy vanish:
\begin{eqnarray}\label{eq6}
\mathcal{B}(N_{\rm E}) = 0, \qquad {E} = \oint \mathcal{B}(N_{\rm E}) = 0.
\end{eqnarray}
This is sensible: a sphere in Minkowski space has vanishing energy, and for a boosted observer it still has vanishing energy.  It should be noted that, because we constrained everything to be axisymmetric, we did not find general boost freedom---only boost freedom along the axis.
Because the Minkowski geometry has this symmetry, naturally our procedure cannot find a unique reference.  Furthermore, because the symmetries of Minkowski are  known we need not look further for any other solutions to our equations to find a good reference.

It is easy to verify that, just as it should, for any $\lambda$ our expression also gives vanishing angular momentum for Minkowski space.

\subsection{Schwarzschild metric}
We next consider the Schwarzschild metric.  To identify the desired physical solution, consider first the limit for small $m$ and/or large $r$; then the physical solution should approach the one just found for Minkowski space.  But expressions of the form (\ref{boostedref}) can only solve the equations (\ref{ELtheta},\ref{ELr}) for the Schwarzschild metric for the trivial value $\lambda=0$.  This is the physical solution we want to use for our preferred reference, both for this limit and for more general values of $m$ and $r$.  With vanishing $T_r,T_\theta$, from (\ref{vector}) for the Schwarzschild metric is that of a static observer:
\begin{eqnarray}\label{eq57}
N^{t}_{\rm E} = 1/\sqrt{-g_{00}}, %\sqrt{\frac{-1}{g_{00}}},
\quad N^{\varphi}_{\rm E} = 0, \quad N^{r}_{\rm E} = 0, \quad N^{\theta}_{\rm E} = 0,
\end{eqnarray}
which is only meaningful for $-g_{00} = 1 - 2m/r > 0$, i.e., outside the horizon.  This strategy cannot be used to find a reference for the dynamic region inside the horizon (alternatives that can cope with the interior region have been discussed elsewhere~\cite{Wu10,Wu11,Wu12}).
For this case the energy boundary term (\ref{optimalenergy}) and the associated energy values are
\begin{eqnarray}
&& \mathcal{B}_{\rm E} = \frac{4r}{16\pi} (1 - \sqrt{1 - {2m}/{r}}) \sin\theta d\theta \wedge d\varphi,
\nonumber\\
\label{eq8}
&& {E} = \oint \mathcal{B}_{\rm E} = r (1 - \sqrt{1 - {2m}/{r}}) = \frac{2m}{1 + \sqrt{1-{2m}/{r}}}.
\end{eqnarray}
This is the well known quasi-local energy first found by Brown and York~\cite{BY93}.

From (\ref{J}) it is easy to see that, as it should, our quasi-local angular momentum vanishes for the Schwarzschild metric.

\subsection{Kerr metric}
Let us now consider specifically the Kerr metric. In terms of the convenient Boyer-Linquist coordinates, the Kerr solution is
\begin{eqnarray}
ds^2 &=& - \frac{(\Delta - a^2 \sin^2\theta)}{\Sigma} dt^2 + \frac{\Sigma}{\Delta} dr^2 + \Sigma d\theta^2
\nonumber\\
&&-\frac{4 m a r \sin^2\theta}{\Sigma} dt d\varphi + \left[ \frac{(r^2 + a^2)^2 - \Delta a^2 \sin^2\theta}{\Sigma} \right] \sin^2\theta d\varphi^2, \label{Kerr}
\end{eqnarray}
where $\Sigma(r, \theta) := r^2 + a^2 \cos^2\theta$, $\Delta(r) := r^2 - 2 m r + a^2$.
For the Kerr metric the special relations $g_{22}/g_{11} = \Delta$ and $g_{03}^2 - g_{00} g_{33} = \Delta\sin^2\theta$ hold.

By the same argument used for the Schwarzschild metric, we take as our desired reference the trivial solution $T_\theta = 0 = T_r$.  The associated displacement vector field from (\ref{vector},\ref{Kerr}) has a well-known form.
For the Kerr metric the expression (\ref{optimalenergy}) is too complicated to find an analytical result.  So we turn to other procedures.  We first find an approximate quasi-local energy for a slowly rotating Kerr black hole, then we present a numerical evaluation.

\subsubsection{When $|a| \ll m$}
With the Kerr metric values the slow rotation approximation is
\begin{eqnarray}\label{eq12}
\mathcal{B}_{\rm E} &\approx& \frac{1}{16\pi} \Biggl\{ 4 r \Bigl( 1 - \sqrt{1 - \frac{2m}{r} + \frac{a^2}{r^2}} \Bigr) \sin\theta
\nonumber\\
&& + \Bigl[ \Bigl( \frac{2 \sin\theta \cos^2\theta}{r} + \frac{2 m \sin^3\theta}{r^2} \Bigr) \sqrt{1 - \frac{2m}{r} + \frac{a^2}{r^2}}
\nonumber\\
\label{eq13}
&& + \frac{2 \sin^3\theta}{r} - \frac{20 m \sin\theta \cos^2\theta}{r^2} + \frac{8 m \sin\theta}{r^2} \Bigr] a^2 \Biggr\} d\theta \wedge d\varphi,
\nonumber\\
\label{eq14}
{E} &=& \oint \mathcal{B}_{\rm E}
\nonumber\\
&\approx& \!\! r \Bigl( 1 - \sqrt{1 - \frac{2m}{r} + \frac{a^2}{r^2}} \Bigr) + \frac{(1 + \frac{2m}{r}) \sqrt{1 - \frac{2m}{r} + \frac{a^2}{r^2}} + 2 + \frac{2m}{r}}{6r} a^2. \qquad
\end{eqnarray}
Our quasi-local energy value for the Kerr metric using the slow rotation approximation turns out to be identical with that found by Martinez from the Brown-York quasi-local expression~\cite{Mart94}.

\subsubsection{General cases: $0\leq|a|\leq m$}
On the $t,r$ constant closed 2-surface we calculated numerically the value of our quasi-local energy with
our best matched reference for the Kerr metric.
In Figure~1 we show how the  quasi-local energy in units of $m$ (i.e., $E/m$) depends on $a/m$ and $r/m$.  Some of the numerical data is presented in Table 1.
\begin{figure}[pb]
\begin{center}
  \includegraphics[trim=20 20 20 20, clip,width=3in,angle=270]{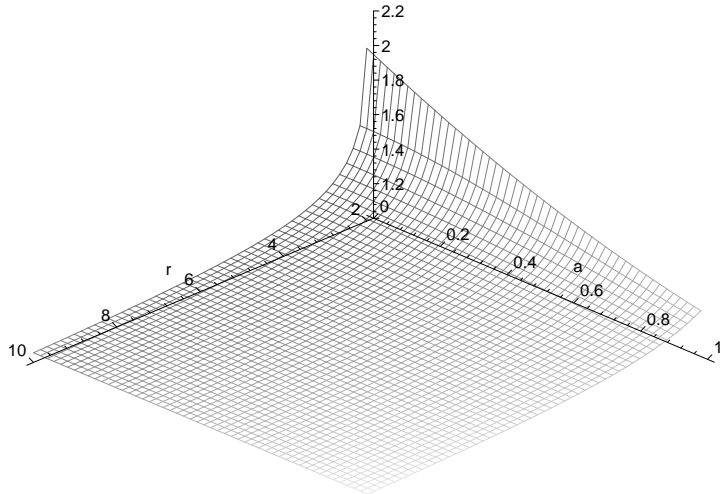}\\
  \vspace*{8pt}
  \caption{A plot of our Kerr quasi-local energy for $r\geq2$, with all the quantities given in units of $m$.  No matter what the value of $a\le 1$, as $r$ approaches infinity, $E$  approaches $1$, as is well known. The highest point corresponds to the Schwarzschild case, with $E=2$ at $r=2$. For any $r$, the larger $|a|$ is, the smaller $E$ is. }
  \end{center}
\end{figure}

\begin{table}
\caption{The quasi-local energy ${E}$ dependance on $|a|$ and  $r$.}
%%%jmn The 100 column has been removed. Furthermore the data was rounded to one less digit---to make the first entry have its known exact analytic value.
{\begin{tabular}{|c|c|c|c|c|c|c|}
\hline
  ${E}/m$&$r/m=2$&$r/m=3$&$r/m=10$&$r/m=20$&$r/m=50$\\\hline
  $|a|/m=0.0$ & 2.00000000 & 1.26794919 & 1.05572809 & 1.0263340 & 1.0102052  \\\hline
  $|a|/m=0.1$ & 1.90258182 & 1.26708057 & 1.05571462 & 1.0263325 & 1.0102050  \\\hline
  $|a|/m=0.2$ & 1.81063676 & 1.26449931 & 1.05567426 & 1.0263278 & 1.0102048  \\\hline
  $|a|/m=0.3$ & 1.72457134 & 1.26027757 & 1.05560706 & 1.0263200 & 1.0102042  \\\hline
  $|a|/m=0.4$ & 1.64468939 & 1.25453045 & 1.05551316 & 1.0263091 & 1.0102036  \\\hline
  $|a|/m=0.5$ & 1.57117096 & 1.24740880 & 1.05539271 & 1.0262951 & 1.0102028  \\\hline
  $|a|/m=0.6$ & 1.50406598 & 1.23909055 & 1.05524594 & 1.0262779 & 1.0102017  \\\hline
  $|a|/m=0.7$ & 1.44330154 & 1.22977111 & 1.05507309 & 1.0262577 & 1.0102004  \\\hline
  $|a|/m=0.8$ & 1.38869944 & 1.21965415 & 1.05487447 & 1.0262344 & 1.0101990  \\\hline
  $|a|/m=0.9$ & 1.33999980 & 1.20894300 & 1.05465044 & 1.0262080 & 1.0101974  \\\hline
  $|a|/m=1.0$ & 1.29688677 & 1.19783368 & 1.05440138 & 1.0261786 & 1.0101956  \\\hline
\end{tabular}}
\end{table}

\subsubsection{Kerr angular momentum}

Evaluating (\ref{J}) explicitly for the Kerr metric (\ref{Kerr}) we find that the exact boundary expression for our quasi-local angular momentum on the $t,r$  constant 2-surface works out to
\begin{eqnarray}
16 \pi \mathcal{B}_{\rm AM} &=& \frac{2 m a \sin^{3}\theta}{\Sigma} [ (1 - 2 r^{2}/\Sigma) (r^{2} + a^{2}) - 2 r^{2} ]\, d\theta \wedge d\varphi
\nonumber\\
&=& - \frac{2 m a [ 3 r^4 + r^2 a^2 (1 + \cos^2\theta) - a^4 \cos^2\theta ] {\sin^3\theta}}{(r^2 + a^2 \cos^2\theta)^2} ~d\theta \wedge d\varphi. \qquad \label{KerrBAM}
\end{eqnarray}
Hence the one non-vanishing component of our quasi-local angular momentum for the Kerr metric is
\begin{eqnarray}
\label{eq11} {J_\varphi} = \oint \mathcal{B}_{\rm AM} = \frac{m a \cos\theta (3 r^2 - r^2 \cos^2\theta + a^2 + a^2 \cos^2\theta)}{4 (r^2 + a^2 \cos^2\theta)} \Big|^{\pi}_{0} = - m a.
\end{eqnarray}

It is amazing that our quasi-local angular momentum turns out to have a simple natural $r$-independent value, even though the boundary term ${\cal B}(\partial_\varphi)$ (\ref{KerrBAM}) is a not so simple function of $r$.

%%%%%%%%%%%%%%%%%%%%%%%%%%%%%%%%%%%%%%%%%%%%%%%%%%%%%%%%%%%%%%%%%%%%%%
\section{Conclusion}
%%%%%%%%%%%%%%%%%%%%%%%%%%%%%%%%%%%%%%%%%%%%%%%%%%%%%%%%%%%%%%%%%%%%%%

Our approach to identifying the quasi-local energy and angular momentum for gravitating systems is via the covariant Hamiltonian formalism.
How to choose the reference geometry and the evolution vector that appear in the Hamiltonian boundary term have been outstanding issues in the quasi-local program. Minkowski geometry and its associated Killing fields are the natural choice, but one actually needs a program for selecting a specific Minkowski geometry on the boundary.
Here for axisymmetric metrics we show that this can be accomplished using an appealing program: \emph{four}-dimensional isometric matching on the boundary 2-surface and energy optimization.

Exact 4D isometric matching includes 2-surface isometric matching.  In general this is the problem of embedding a closed 2-surface into Minkowski space.  Wang and Yau have carefully considered this problem in connection with their quasi-local energy investigations~\cite{WY09PRL,WY09CMP}.  The problem generally involves the solution of a nonlinear partial differential equation.  Although they proved an existence and uniqueness theorem, it does not seem possible to find a general analytic formula for the embedding functions.  However, for the axisymmetric case  considered here the situation is quite different.
The 2-surface isometric matching, and indeed the whole 4D isometric matching is in this case essentially a simple \emph{algebraic} problem. With 4D isometric matching the choice of Minkowski reference on the boundary still has two degrees of freedom.

To determine the remaining two degrees of freedom we have proposed energy optimization.
Our energy optimization procedure in the axisymmetric case leads to a pair of first order quasi-linear \emph{ordinary\/} differential equations.
A certain class of Kerr-like metrics fortunately have a simple analytic solution (the trivial solution) to these non-linear energy optimization equations.  One cannot expect to find such simple analytic solutions in more general cases, such as a dynamical metric with more non-vanishing cross terms, a time dependent metric, or for some other 2-surface $S$ described by a formula like $R = R(r,\theta) = \hbox{constant}$.  That we can have our solution of trivial embedding variables can be attributable to the fact that the Boyer-Lindquist coordinate $r$ has a special significance for these Kerr-like geometries, and the constant $r$ surfaces are thus quite special and meaningful.

For the considered metrics it is noteworthy that our optimal timelike vector field turned out to be hypersurface orthogonal, corresponding to a static observer. Our quasi-local expression with  best matched reference gives for Minkowski space, as expected, vanishing energy and angular momentum.  For the Schwarzschild metric it gives the well-known energy value first found by Brown and York along with vanishing angular momentum.

We found the exact analytic formula for our optimal quasi-local energy boundary term for these Kerr-like metrics on the surface $r=\hbox{constant}$.
For the Kerr metric in the slow rotation approximation our quasi-local energy agrees with that  found by Martinez using the Brown-York quasi-local expression.
We plotted our Kerr quasi-local energy and note that it is a decreasing function of $|a|$.  At present we have no physical understanding as to why this is the case.

For these Kerr-like metrics we found our optimal quasi-local angular momentum boundary term.  Our quasi-local angular momentum for the Kerr metric worked out to be an unexpectedly simple but satisfying constant value ($-ma$).
This may again be attributed to the chosen 2-surface $S$, which is given by a constant value of the Boyer-Lindquist coordinate $r$.  It is another indication of the special significance of this coordinate for these metrics.  Our quasi-local angular momentum would not be so simple for closed surfaces of the form $R=R(r,\theta)=\hbox{constant}$.

Encouraged by these results, the application of these ideas is being extended to both general axisymmetric metrics and completely general metrics.

%%%%%%%%%%%%%%%%%%%%%%%%%%%%%%%%%%%%%%%%%%%%%%%%%%%%%%%%%%%%%%%%%%%%%%
\section*{Acknowledgments}
%%%%%%%%%%%%%%%%%%%%%%%%%%%%%%%%%%%%%%%%%%%%%%%%%%%%%%%%%%%%%%%%%%%%%%
This work was supported by the National Science Council of the R.O.C. under the grants NSC 101-2112-M-008-006 and 99-2112-M-008-005-MY3, the Institute of Physics, Academia Sinica, and the National Center of Theoretical Sciences (NCTS).

%%%%%%%%%%%%%%%%%%%%%%%%%%%%%%%%%%%%%%%%%%%%%%%%%%%%%%%%%%%%%%%%%%%%%%
%\section*{References}
%%%%%%%%%%%%%%%%%%%%%%%%%%%%%%%%%%%%%%%%%%%%%%%%%%%%%%%%%%%%%%%%%%%%%%

\end{document}